\documentclass[conference]{IEEEtran}
\IEEEoverridecommandlockouts
\usepackage{cite}
\usepackage{amsmath,amssymb,amsfonts}
\usepackage{graphicx}
\usepackage{textcomp}
\usepackage{subcaption}
\usepackage{soul}
\usepackage{algorithm}
\usepackage[noend]{algpseudocode}
\usepackage{multirow}
\usepackage{array}
\usepackage{amsmath}
\usepackage[font=small,skip=0.001cm]{caption}
\usepackage{tabularx}
\usepackage{float}
\usepackage{listings}
\usepackage{xcolor}
\def\BibTeX{{\rm B\kern-.05em{\sc i\kern-.025em b}\kern-.08em
    T\kern-.1667em\lower.7ex\hbox{E}\kern-.125emX}}
\begin{document}

\title{Capability-Based Multi-Tenant Access Management in Crowdsourced Drone Services
}

\author{
    \IEEEauthorblockN{Junaid Akram\IEEEauthorrefmark{1}, Ali Anaissi\IEEEauthorrefmark{1}\IEEEauthorrefmark{2}, 
    Awais Akram\IEEEauthorrefmark{3}, Youcef Djenouri\IEEEauthorrefmark{4}\IEEEauthorrefmark{5}\thanks{Corresponding Authors: Youcef Djenouri, Rutvij H. Jhaveri}, 
    Palash Ingle\IEEEauthorrefmark{6}, Rutvij H. Jhaveri\IEEEauthorrefmark{7}}
    
    \IEEEauthorblockA{\IEEEauthorrefmark{1}School of Computer Science, The University of Sydney, Camperdown NSW 2008, Australia\\
    \IEEEauthorrefmark{2}TD School, University of Technology Sydney, Ultimo NSW 2007, Australia\\
    \IEEEauthorrefmark{3}Independent Researcher\\
    \IEEEauthorrefmark{4}NORCE Norwegian Research Center, Oslo, Norway\\
    \IEEEauthorrefmark{5}University of South-Eastern Norway, Konsberg, Norway\\
    \IEEEauthorrefmark{6}Department of Computer and Information Security, Sejong University, Seoul, South Korea\\
    \IEEEauthorrefmark{7}Department of Computer Science and Engineering, School of Technology, Pandit Deendayal Energy University, India\\
    Email: jakr7229@uni.sydney.edu.au, ali.anaissi@sydney.edu.au, awais.akram.1212@gmail.com, \\ yodj@norceresearch.no, palash@sejong.ac.kr, rutvij.jhaveri@sot.pdpu.ac.in}
}

\maketitle

\begin{abstract}
We propose a capability-based access control method that leverages OAuth 2.0 and Verifiable Credentials (VCs) to share resources in crowdsourced drone services. VCs securely encode claims about entities, offering flexibility. However, standardized protocols for VCs are lacking, limiting their adoption. To address this, we integrate VCs into OAuth 2.0, creating a novel access token. This token encapsulates VCs using JSON Web Tokens (JWT) and employs JWT-based methods for proof of possession. Our method streamlines VC verification with JSON Web Signatures (JWS) requires only minor adjustments to current OAuth 2.0 systems. Furthermore, in order to increase security and efficiency in multi-tenant environments, we provide a novel protocol for VC creation that makes use of the OAuth 2.0 client credentials grant. Using VCs as access tokens enhances OAuth 2.0, supporting long-term use and efficient data management. This system aids bushfire management authorities by ensuring high availability, enhanced privacy, and improved data portability. It supports multi-tenancy, allowing drone operators to control data access policies in a decentralized environment.
\end{abstract}


\section{Introduction}

Unmanned Aerial Vehicles (UAVs), or drones, have transformed various sectors, including surveillance, logistics, and environmental monitoring, over the last two decades\cite{10547221}. More recently, the concept of crowdsourced drone services within the Internet of Drone Things (IoDT) has enabled a decentralized and collaborative approach to environmental management, with significant applications in critical areas like bushfire monitoring\cite{akram2024priv,akram2024d2xchain}. This paper explores the deployment of a crowdsourced drone service framework specifically for bushfire management, where authorities act as data consumers and drone operators provide real-time data for decision-making\cite{akram2024DroneSSL,10535995}. This model democratizes drone usage, allowing more scalable and rapid responses to bushfire events through the aggregation of critical data from multiple sources\cite{10492460,akram2024ddrm}.

However, the implementation of such a decentralized system introduces several challenges. The most prominent of these is managing access control in a multi-tenant environment, where numerous independent drone operators must handle access rights while maintaining security and operational efficiency\cite{akram2022bc}. Traditional access control mechanisms, such as bearer tokens and static Access Control Lists (ACLs), are often inadequate for these dynamic environments. In scenarios where resource servers operate independently of authorization servers ($S_{auth}$) or where drones need to function offline or with intermittent connectivity, these methods fail to provide the required flexibility and security\cite{chadwick2019improved,10535995}.

The core challenge lies in designing an access management system that can dynamically handle the decentralized nature of drone services. During bushfire emergencies, for example, drone operators need to provide timely data access without the delays caused by conventional static systems. Furthermore, the crowdsourced nature of the system raises concerns about data privacy and the quality of information provided by different operators, making it crucial to secure both the access and the integrity of the shared data\cite{akram2024priv}.

To overcome these limitations, we propose a capability-based access control system that integrates OAuth 2.0 with Verifiable Credentials (VCs). OAuth 2.0, typically used for web authorization, is adapted here to manage the dynamic access requirements of drone services, while VCs provide a secure and privacy-preserving way to verify access rights\cite{ace_oauth_authz}. This approach addresses the weaknesses of traditional methods, offering a flexible solution for decentralized, ad-hoc access requests in complex environments like bushfire management, ensuring secure, reliable data exchange between operators and authorities.
Our work makes the following key contributions:

\begin{itemize}
\item We create a protocol based on OAuth 2.0 for issuing VCs, optimizing the OAuth framework for decentralized drone operations and enhancing security and flexibility in multi-tenant environments.
\item We enhance JSON Web Tokens (JWTs) by integrating VCs, enabling secure, standardized, and easily verifiable data exchanges between drone operators and bushfire management authorities.
\item We propose improvements to OAuth 2.0 workflows with a streamlined proof-of-possession mechanism for secret keys, simplifying verification to a single message exchange, thus reducing complexity and bolstering security.
\item We design a cloud storage system optimized for drone-operated bushfire detection, ensuring lightweight, easily integrated infrastructure to facilitate efficient data management and seamless adoption.
\end{itemize}

The paper is organized as follows: Section II covers related work, Section III outlines the proposed solution's architecture, and Section IV details the system design. Section V presents performance and security evaluations, and Section VI concludes with key contributions and future directions.

\section{Related Work}

With an emphasis on IoT devices, Lagutin et al. \cite{lagutin2019enabling} propose incorporating VCs and DIDs into OAuth 2.0 through the use of ACE-OAuth \cite{ace_oauth_authz}, a lightweight variant of OAuth 2.0 made for constrained environments. Their solution employs DIDs and VCs  as authentication grants to obtain access tokens, whereas our approach uses authentication grants to issue VCs as access tokens. Other innovative approaches, such as DIDComm \cite{didcomm_messaging}, Presentation Exchange \cite{presentation_exchange}, and Hyperledger Aries \cite{hyperledger_aries}, develop protocols for secure communication and credential management but lack compatibility with existing authorization standards. 
Munoz's \cite{eidas_ssi} investigation on integrating DIDs and VCs into eIDAS is one example of efforts to incorporate these technologies into official systems of identification . Our solution handles delegation and attenuation while utilizing VCs to convey capabilities and taking use of their standard format. These concepts are challenging to implement solely with VCs. Macaroons\cite{birgisson2014macaroons} and ZCAP-LD\cite{zcap_spec} offer alternative methods for capability-based systems, considered in our design.
By addressing the limitations of existing approaches and focusing on the unique requirements of crowdsourced drone services for bushfire management, our work advances the state-of-the-art in capability-based access control and multi-tenant resource management.

\section{System Design}
\subsection{Entities and Definitions}

\begin{figure}[t]
    \centering
    \includegraphics[width=\columnwidth]{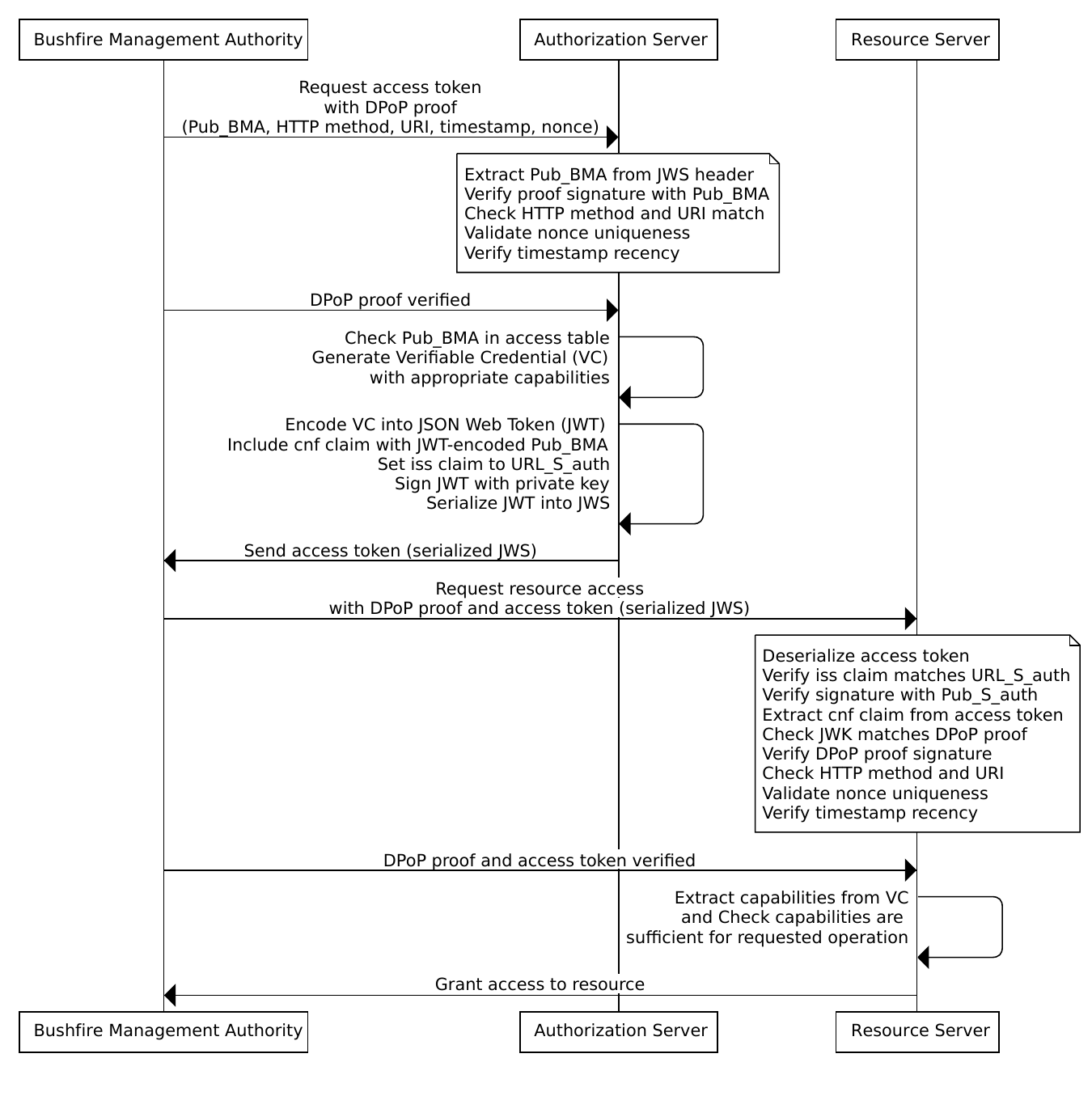}
    \caption{Capability-Based Multi-Tenant Access Management Framework}
    \label{fig:Multitenant}
\end{figure}

In this section, we outline the key entities and definitions integral to our capability-based multi-tenant access management system, which is tailored for crowdsourced drone services in bushfire management. The framework is depicted in Figure \ref{fig:Multitenant}. Our framework includes multiple drone operators as data providers, who capture and deliver real-time information essential for bushfire monitoring and response. Bushfire management authorities serve as data consumers, utilizing this information to enhance decision-making and emergency response strategies. Each drone operator manages resources, such as drone-collected data stored on a shared Resource Server ($S_{res}$), and maintains an $S_{auth}$ to handle access rights. The primary goal of our system is to enable secure and efficient access to this valuable data for bushfire management authorities through access tokens generated by the $S_{auth}$.

The entities are defined as follows: a Bushfire Management Authority (BMA) is identified by a unique public key \( \text{Pub}_{\text{BMA}} \) and requests access to data. An $S_{auth}$ is identified by a unique URL (\( URL_{S_{auth}} \)) and public key \( \text{Pub}_{\text{$S_{auth}$}} \). The $S_{auth}$ generates and manages access tokens. A $S_{res}$ stores drone-collected data and enforces access control policies as specified by the $S_{auth}$. Capabilities, denoted as \( C_{\text{op} \rightarrow \text{res}} \), define specific operations, such as allowing a BMA to read data from a sensor.

The system operates as follows: A BMA requests an access token from the $S_{auth}$ using its public key \( \text{Pub}_{\text{BMA}} \). The $S_{auth}$ issues an access token if the BMA holds a valid Verifiable Credential (VC) that includes the required capabilities. The $S_{auth}$ maintains an access table mapping each BMA's public key to its capabilities, \( \text{Pub}_{\text{BMA}} \rightarrow [C_1, C_2, \ldots, C_n] \). The $S_{res}$ maintains a resource table linking resource identifiers to their respective $S_{auth}$ URLs and public keys, ensuring accurate access control delegation. To ensure secure and consistent access, each $S_{res}$ adheres to a standardized credential format, promoting interoperability across the system.

\subsection{Proof-of-Possession of a Key}

Our system employs \textit{DPoP} \cite{fett2020oauth} to validate cryptographic key possession. While DPoP is currently a draft with the IETF, it has gained significant attention and is actively being developed. Future iterations of our system might incorporate additional proof-of-possession techniques. DPoP, designed for HTTP communications, facilitates proof-of-possession through a single message. In our implementation, drone operators (acting as data providers) embed a DPoP proof within their HTTP requests to bushfire management authorities (acting as clients). A JSON Web Signature (JWS) signed using the operator's private key constitutes a DPoP proof. The JWS header provides a public key for the JSON Web Key (JWK) that may be used to verify the signature of the DPoP proof, along with a type field that is set to \textit{dpop+jwt} and the digital signature method. The JWS payload consists of the request's HTTP method, HTTP URI, creation timestamp, and unique identifier (such as an adequately large random integer).
Listing \ref{lst:dpop_proof} offers an illustration of a DPoP proof that is employed in our framework. The JWS payload is shown on lines 10-15, whereas the JWS header is shown on lines 1-9. Lines 4–7 include the public key that is required to validate the DPoP proof's digital signature. An HTTP POST request to "https://drone-services.org/token" includes this DPoP evidence.

\begin{lstlisting}[caption={Example of a DPoP proof for drone data access}, label={lst:dpop_proof}]
{
  "typ": "dpop+jwt",
  "alg": "EdDSA",
  "jwk": {
    "kty": "OKP",
    "crv": "Ed25519",
    "x": "3pLJ...sXIS7"
  }
}
{
  "htm": "POST",
  "htu": "https://drone-services.org/token",
  "iat": 1617548847,
  "jti": "a1d2e4...gcd567"
}
\end{lstlisting}

In this example, the JWS header specifies the use of the EdDSA algorithm (\texttt{"EdDSA"}) and includes the JWK containing the OKP public key. The request URI, the issued-at timestamp, the HTTP method ("POST"), and a unique identifier (JWT ID) are all included in the payload. This proof mechanism ensures secure and verifiable communication between drone operators and bushfire management authorities, guaranteeing that data exchange is authenticated and authorized.

\subsection{Access Token Request}

The BMA request access token from the $S_{auth}$ in accordance with our protocol. DPoP proof with signature verifiable using \( \text{Pub}_{\text{BMA}} \) is attached to this request. In order to verify the DPoP proof, $S_{auth}$ does the following actions: It first extracts \( \text{Pub}_{\text{BMA}} \) from the JWS header, then uses \( \text{Pub}_{\text{BMA}} \) to verify the proof's signature. It then makes sure the HTTP method and URI in the payload match the request made by the BMA, validates that the random number in the payload is unique, and determines the proof's recency based on when it was created.
When the proof is successfully validated, $S_{auth}$ recognizes \( \text{Pub}_{\text{BMA}} \) in its access table. After that, it creates a VC that is in line with the Resource Server's ($S_{res}$) credential description and embeds all pertinent capabilities associated with \( \text{Pub}_{\text{BMA}} \). Following that, this VC is encoded in accordance with the OAuth 2.0 protocol to create a JSON Web Token (JWT). The JWT includes the \texttt{cnf} (confirmation) claim with the JWT-encoded \( \text{Pub}_{\text{BMA}} \) and the \texttt{iss} (issuer) claim set to \( \text{URL}_{S_{auth}} \). Finally, $S_{auth}$ encodes the JWT into a JWS, signs it using its private key, and serializes it with \textit{base64url} encoding. This serialized JWS, now functioning as an access token, is returned to the BMA.

\subsection{Resource Request} \label{sec:resource_request}

The system handles the request to perform an action on the resource when a BMA provides a DPoP proof and its acquired access token. More complicated cases requiring multiple VCs are covered in Section \ref{sec:combined_vcs}. Upon receiving a request, the $S_{res}$ performs the following verifications: it opens the resource table corresponding to the requested resource, extracts \( \text{URL}_{S_{auth}} \) and \( \text{Pub}_{S_{auth}} \), deserializes the access token, verifies that the \texttt{iss} claim matches \( \text{URL}_{S_{auth}} \), and uses \( \text{Pub}_{S_{auth}} \) to verify the signature. 
After that, it takes the deserialized access token and extracts the \texttt{cnf} claim, ensuring that it has the same JWK as the DPoP proof. Next, it confirms the signatures of the DPoP proof, looks up the HTTP URL and method, makes sure the identification is unique, and verifies the creation time.
The VC is bound to the BMA that submitted the request via this verification procedure. The $S_{res}$ extracts the capabilities from the VC and verifies they are enough for the proposed operation if all of these steps are completed successfully.

\subsection{Token Revocation}

Our system offers two methods for determining the state of an access token: verifying the included Verifiable Credential (VC) revocation status or using OAuth 2.0 token introspection. In OAuth 2.0 introspection, $S_{res}$ queries the token introspection endpoint provided by $S_{auth}$, retrieves related meta-data, and assesses the token's state. The introspection endpoint returns a JSON object with an \texttt{active} field that indicates the token’s validity. However, this approach increases communication overhead, as $S_{res}$ must inquire about each token separately, and it can compromise client privacy due to the implicit sharing of the introspection endpoint. The alternative method involves using a privacy-preserving revocation mechanism, where $S_{res}$ verifies the access token by checking the included VC's revocation status. Each VC granted by $S_{auth}$ is associated with a position in a revocation list, represented as a bitstring. When a VC is revoked, its corresponding bit is set to 1, and the position is indicated by the \texttt{revocationListIndex} field. $S_{res}$ downloads the revocation list once and can reuse it for multiple VCs, enhancing efficiency and preserving client privacy, as $S_{auth}$ does not know which VC is being verified. Additionally, the VC may provide a URL for retrieving the revocation list, offering flexibility by enabling storage outside of $\text{URL}_{S_{auth}}$. This method is more flexible in handling revocation data than OAuth 2.0 introspection.

\subsection{Combining Multiple VCs} \label{sec:combined_vcs}

The capacity to integrate several VCs into a single Verifiable Presentation (VP) is a significant advantage of VCs. A BMA in our system has the ability to combine numerous access tokens ($S_{auth}$) that it receives from various Authorization Servers into a single access token. The same \( \text{Pub}_{\text{BMA}} \) must appear in the \texttt{cnf} field for every access token.

The following stages are involved in the process of generating a VP:

\begin{enumerate}
    \item A new JWT object is created by the BMA.
    \item Sets \( \text{Pub}_{\text{BMA}} \) as the SHA-256 hash in the \texttt{iss} (issuer) field of this JWT.
    \item Includes an array of each unique access token in a \texttt{vp} field in the JWT.
    \item Encodes the JWT in a JWS.
    \item Signs it with the BMA's private key.
    \item Generates the new access token by serializing the JWS.

\end{enumerate}

The verification process for the $S_{res}$ is as follows:
\begin{enumerate}
    \item Deserializes the access token and checks for a \texttt{vp} field.
    \item It extracts each individual access token in the event that a \texttt{vp} field is present.
    \item Uses the process described in Section \ref{sec:resource_request} to verify each access token.
    \item Uses \( \text{Pub}_{\text{BMA}} \) to confirm the signature of the access token that was received.
    \item It retrieves all capabilities from each VC and verifies that they are sufficient to authorize the intended operation if all verifications are successful.
\end{enumerate}

This approach enables BMAs to combine multiple access tokens into a single, verifiable token. This is particularly useful in a crowdsourced drone services environment where data from various drone operators needs to be accessed and combined to provide a comprehensive and timely response during bushfire emergencies. By verifying all combined access tokens, $S_{res}$ ensures proper authorization of all capabilities, enhancing the security and efficiency of the data management process.

\subsection{Using DIDs as VC Subject}

While our approach employs public keys, standard VC systems use Decentralized Identifiers (DIDs) to identify the credential subject. Key rotation is an important DID feature that our system does not provide. By refreshing the public key linked to a DID, key rotation enhances security and adaptability. An entity often requests a DID registry in order to obtain the public key associated with a DID.
For example, an access token with a DID is first obtained by a BMA through their central command system. However, for the $S_{res}$ to validate the token, it must query the DID registry to get the corresponding public key, used as \( \text{Pub}_{\text{BMA}} \).
Subsequently, the authority decides to use a mobile command unit to access the $S_{res}$. They update their DID to link it to a public key stored on the mobile unit through the DID registry. The $S_{res}$ will retrieve a different \( \text{Pub}_{\text{BMA}} \) from the DID registry when the authority requests resource access again with the same access token.

This mechanism offers several advantages. Enhanced security is achieved by rotating keys, which reduces the risk associated with long-term key exposure. Flexibility is ensured as authorities can switch devices while maintaining the same DID, allowing continuous access without reissuing credentials. Scalability is supported by the system, as it can accommodate a broader range of clients and devices without managing static public keys. Integrating DIDs into our system could significantly improve its robustness and adaptability, especially in the dynamic and decentralized environment of crowdsourced drone services for bushfire management. This would allow drone operators and bushfire management authorities to leverage advanced security features and ensure seamless access across different devices and contexts.

\section{Implementation and Evaluation}

Our capability-based multi-tenant access management system has a proof-of-concept prototype that we have created with Python 3. The JWS functionality is provided by the JWCrypto library, and for cryptographic operations, we employ the EdDSA digital signature technique \cite{josefsson2017edwards} using Ed25519 public keys.

\begin{figure*}[t]
    \centering
    \includegraphics[width=\textwidth]{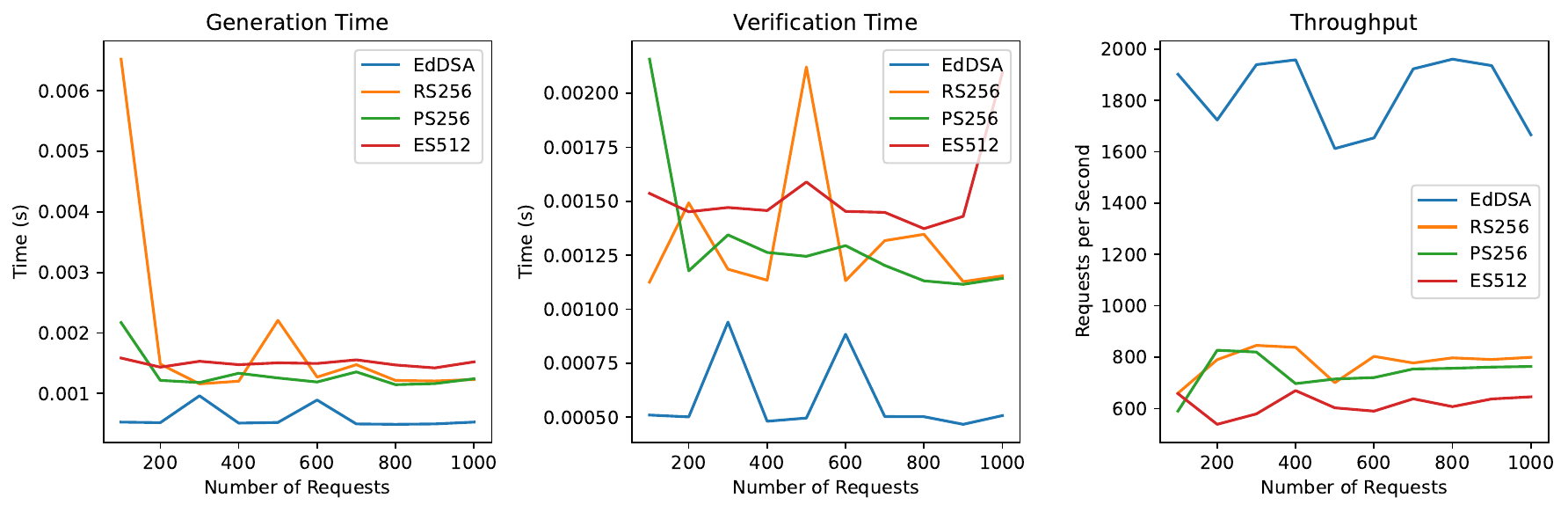}
\caption{Performance comparison of cryptographic algorithms across varying request loads: (a) access token generation time (b) verification time (c) system throughput.}
    \label{fig:performance_comparison}
\end{figure*}

\subsection{Performance analysis}

We developed a prototype for our proposed system as a proof of concept. The system integrates cloud storage where drone operators store their data. Each operator manages an Authorization Server (\(S_{auth}\)), which grants access rights to bushfire management authorities. The public keys of the \(S_{auth}\) in charge of those areas are mapped to storage locations via the resource table in the cloud storage. For instance, the \(S_{auth}\) identified by \( \text{Pub}_{\text{drone1}} \) controls access to the /data/drone1 directory.
Our system defines a VC type that specifies the format for acceptable VC claims. An example VC in our system includes claims formatted as follows:

\begin{lstlisting}[caption={VC Claims Format}, label={lst:vc_format}]
{
  "capabilities": [
    {"/data/drone1": ["read", "write"]},
    {"/data/drone2": ["read"]}
  ]
}
\end{lstlisting}

A VC must contain a \texttt{capabilities} key, which lists paths and their corresponding access rights. Each \(S_{auth}\) maintains an access table that maps a BMA’s public key to its access rights. For example, the \(S_{auth}\) for drone1 includes the public keys of several authorities, specifying their read and write access to various data paths.
An access token is granted to a BMA upon initiation of the access token request protocol. The BMA then sends a base64url-encoded proof-of-possession along with an HTTPS POST requesting resource access to the \(S_{auth}\) token providing service. In response, the \(S_{auth}\) provides a signed access token. The size of claims included affects the size of the access token; a JWS access token normally has 700 bytes. With the access token and an extra HTTP header containing the proof-of-possession, the resource access request is made via HTTPS GET.
The cloud storage goes through many validation processes. For the /data/drone1 path, it retrieves the \(S_{auth}\) public key. Then, it verifies the VC and proof-of-possession. Finally, it checks if the VC capabilities grant the authority read access to the requested data. This end-to-end evaluation shows that our system is practical for secure and efficient data access in a crowdsourced drone services environment.

To comprehensively assess our system's performance, we conducted simulations focusing on access token generation time, verification time, and system throughput for four cryptographic algorithms: EdDSA, RS256, PS256, and ES512, across request loads ranging from 100 to 1000. The results are illustrated in Figure \ref{fig:performance_comparison}. EdDSA demonstrated the fastest access token generation time, consistently averaging 0.58 milliseconds, with a verification time of approximately 0.17 milliseconds. The throughput for EdDSA was notably high, reaching about 5700 requests per second at a load of 1000, indicating its efficiency and suitability for real-time applications.
RS256 showed robust performance with an average generation time of 1.2 milliseconds and a verification time of 0.35 milliseconds. Its throughput was around 3300 requests per second at maximum load, providing a good balance between security and performance. PS256 exhibited similar performance to RS256, with a generation time of 1.1 milliseconds and a verification time of 0.32 milliseconds, achieving approximately 3400 requests per second. These results suggest PS256 as a marginally more efficient option while maintaining comparable security properties.
ES512, the most computationally intensive algorithm, had the longest generation and verification times of 2.3 milliseconds and 0.65 milliseconds, respectively, resulting in the lowest throughput of around 1500 requests per second at the highest load. Despite its strong security guarantees, the significant computational overhead of ES512 may impact its suitability for high-throughput scenarios. These simulations indicate that EdDSA offers the best performance in terms of speed and throughput, making it highly suitable for real-time operations. Both RS256 and PS256 provide a reasonable trade-off between security and performance, while ES512, despite its computational expense, offers robust security features beneficial in environments where security is paramount. 

\subsection{Formal Security Analysis}

In this section, we present the formal security analysis of our proposed solution using the Alloy Analyzer \cite{jackson2019alloy}. Formal methods provide a rigorous foundation for verifying the correctness and security properties of complex systems. Our comprehensive Alloy model encapsulates the key components and behaviors of our system, ensuring robust security properties.

Our formal model describes the key entities and their relationships. The primary entities include Tokens (\(T\)), Authorities (\(A\)), Servers (\(S\)), and Revocation Servers (\(RS\)). Each entity has specific attributes and interacts with other entities in defined ways. Tokens (\(T\)) represent a set of tokens, each with a validity attribute (\(T.\text{valid} \in \{\text{true}, \text{false}\}\)) and an owner attribute (\(T.\text{owner} \in \text{Authority}\)). Authorities (\(A\)) are sets of authorities, each owning a subset of tokens (\(A.\text{tokens} \subseteq T\)). Servers (\(S\)) are sets of servers, each authorizing a subset of tokens (\(S.\text{authorizedTokens} \subseteq T\)). Revocation Servers (\(RS\)) are a subset of servers that handle revocation, maintaining a list of revoked tokens (\(RS.\text{revokedTokens} \subseteq T\)).

To ensure the security of our system, we define several predicates and assertions that capture the desired security properties. The valid token usage property ensures that any access token used by a BMA is valid. The holder of the token must possess the appropriate VCs issued by an $S_{auth}$, granting them the necessary capabilities. The model confirms that the resource server correctly validates these tokens against the holder’s capabilities before granting access to the requested data. This is formally expressed as:

\begin{eqnarray*}
\begin{aligned}
\text{validTokenUsage}(a, t) \iff & \, t \in A[a].\text{tokens} \land \\
                                  & \, t.\text{valid} = \text{true}
\end{aligned}
\end{eqnarray*}

To further ensure security, we assert that no tampered tokens are used. This assertion checks that for all authorities and tokens, if a token is valid, it must belong to the authority and be valid:
\begin{eqnarray*}
\forall a \in A, t \in T \cdot (t.\text{valid} \implies \text{validTokenUsage}(a, t))
\end{eqnarray*}

The token integrity property ensures that any token used by a server is valid. The server must detect and reject any forged or tampered tokens. This is formally expressed as:
\begin{eqnarray*}
\text{detectForgedTokens}(s, t) & \iff & t \in \\ S[s].\text{authorizedTokens} \land  t.\text{valid} = \text{true}
\end{eqnarray*}

We assert that no forged tokens are used, ensuring that for all servers and tokens, the server must detect and reject invalid tokens:
\begin{eqnarray*}
\forall s \in S, t \in T \cdot \text{detectForgedTokens}(s, t)
\end{eqnarray*}

The revocation mechanisms property ensures that any attempt to use a revoked credential is denied, maintaining the integrity of access control. This is crucial for mitigating the risks associated with compromised or outdated credentials. This is formally expressed as:

\begin{eqnarray*}
\text{tokenRevocationMechanism}(rs, t) & \iff & \\ t \notin RS[rs].\text{revokedTokens}
\end{eqnarray*}

We assert that revocation mechanisms are effective, ensuring that for all revocation servers and tokens, the revocation mechanism must effectively revoke tokens:
\begin{eqnarray*}
\forall rs \in RS, t \in T \cdot \text{tokenRevocationMechanism}(rs, t)
\end{eqnarray*}

Our system supports the principles of delegation (transferring capabilities) and attenuation (limiting capabilities). For instance, a client can delegate an access token to another client while limiting access to only a subset of the original capabilities. This ensures that the system can flexibly manage access rights in a secure manner.

\section{Conclusions}

This study proposed a capability-based access control system for crowdsourced drone services in bushfire management, securing resources in a multi-tenant environment. Verifiable Credentials (VCs) were used to represent user capabilities, integrated with OAuth 2.0 for requesting and utilizing these credentials. JSON Web Tokens (JWT) and JSON Web Signatures (JWS) facilitated smooth integration, leveraging existing infrastructure. Public keys were chosen over DIDs for simplicity, though DIDs can be integrated if needed. This enhances OAuth 2.0 for long-term credential use and efficient data management. Our solution provides high availability, enhanced privacy, data portability, and multi-tenant support, empowering drone operators to control access in a decentralized setting, crucial for bushfire emergencies.

 \bibliographystyle{ieeetr}
 \bibliography{ref}

\end{document}